\documentclass[aps,prx,floatfix,superscriptaddress,showpacs,twocolumn,nofootinbib,longbibliography]{revtex4-1}

\usepackage{graphicx}
\usepackage{amsmath}
\usepackage{amssymb}
\usepackage{amsfonts}
\usepackage{amsthm}
\usepackage{mathtools}
\usepackage{siunitx}
\usepackage[unicode=true, breaklinks=true, pdfborder={0 0 1}, backref=false, colorlinks=true, linkcolor=blue, citecolor=blue, urlcolor=blue]{hyperref}
\usepackage{color}
\usepackage{soul}
\usepackage{textcomp}
\usepackage{bm}
\usepackage{dsfont}
\usepackage{relsize}
\usepackage{braket}
\usepackage{enumitem}   
\usepackage{mathrsfs}
\usepackage{cleveref}
\usepackage{bigints}
\usepackage{framed}
\usepackage[makeroom]{cancel}
\usepackage{comment}

\theoremstyle{plain}

\numberwithin{obs}{section}

\definecolor{Blue}{rgb}{0,0,1}
\definecolor{Red}{rgb}{1,0,0}
\definecolor{Green}{rgb}{0,1,0}
\definecolor{darkgreen}{rgb}{0,.7,0}
\definecolor{Purp}{rgb}{.2,0,.2}
\definecolor{white}{rgb}{1,1,1}

\begin{document}

\title{ 
Experimental verification of the work fluctuation-dissipation relation for information-to-work conversion 
}

\author{David Barker}
\affiliation{NanoLund and Solid State Physics, Lund University, Box 118, 22100 Lund, Sweden}
\author{Matteo Scandi}
\affiliation{ICFO - Institut de Ciencies Fotoniques, The Barcelona Institute of Science and Technology, Castelldefels (Barcelona), 08860, Spain}
\author{Sebastian Lehmann}
\affiliation{NanoLund and Solid State Physics, Lund University, Box 118, 22100 Lund, Sweden}
\author{Claes Thelander}
\affiliation{NanoLund and Solid State Physics, Lund University, Box 118, 22100 Lund, Sweden}
\author{Kimberly A. Dick}
\affiliation{NanoLund and Solid State Physics, Lund University, Box 118, 22100 Lund, Sweden}
\affiliation{Centre for Analysis and Synthesis, Lund University, Box 124, 22100 Lund, Sweden.}
\author{Mart\'{i} Perarnau-Llobet}
\affiliation{D\'{e}partement de Physique Appliqu\'{e}e, Universit\'{e} de Gen\'{e}ve, Gen\'{e}ve, Switzerland}
\author{Ville F. Maisi}
\affiliation{NanoLund and Solid State Physics, Lund University, Box 118, 22100 Lund, Sweden}

\date{\today}

\begin{abstract}
We study experimentally work fluctuations in a Szilard engine that extracts work from information encoded as the occupancy of an electron level in a semiconductor quantum dot. We show that as the average work extracted per bit of information increases towards the Landauer limit $k_BT \ln 2$, the work fluctuations decrease in accordance with the work fluctuation-dissipation relation. 
We compare the results to a protocol without measurement and feedback and show that when no information is used, the work output and fluctuations vanish simultaneously contrasting the information-to-energy conversion case where increasing amount of work is produced with decreasing fluctuations. Our work highlights the importance of fluctuations in the design of information-to-work conversion processes. 

\end{abstract}

\maketitle

When considering thermodynamic processes at the microscale, where thermal and quantum fluctuations play a dominant role,   one faces the challenge of describing
work and heat as stochastic quantities. 
Understanding and characterising their statistics  has become an  active area of research, and is one of the main aims of the fields of  stochastic and quantum thermodynamics \cite{Jarzynski2011,Seifert_2012,Esposito2009,Campisi2011c,Goold_2016}. 
The exact form of heat and work distributions  highly depends on the specific protocol in question.  
The situation considerably simplifies when considering finite-time processes where the system  remains close to the instantaneous equilibrium state (i.e., in the linear-response regime with respect to the driving speed) \cite{nultonQuasistaticProcessesStep1985b,speckDistributionWorkIsothermal2004,Crooks2007,hoppenauWorkDistributionQuasistatic2013a,Kwon2013,Bonana2014,Mandal2016a,Scandi2020quantum,Bhandari2020Geometric}.
In this case, and when no quantum coherence is generated along the process, the work distribution takes a Gaussian shape  \cite{speckDistributionWorkIsothermal2004,hoppenauWorkDistributionQuasistatic2013a,Scandi2020quantum}, with its first and second moments related through the work fluctuation-dissipation relation (FDR)~\cite{hermans_simple_1991,jarzynski_nonequilibrium_1997,Mandal2016,Miller2019work}:
\begin{equation}
    W_{diss} = \frac{1}{2k_BT}\sigma^2_W.
    \label{eq:FDR}
\end{equation}
Here, $T$ is the temperature of the surrounding environment, $k_B$ is Boltzmann's constant,  $W_{\rm diss}$ is the average dissipated work along the process: 
\begin{equation}
    W_{diss} :=  \langle W \rangle-\Delta F ,
    \label{eq:Wdiss}
\end{equation}
i.e., the difference between the mean work~$\langle W \rangle$ performed  to the system and the change in equilibrium free energy~$\Delta F$, and finally~$\sigma_W^2$ is the variance of the work distribution. The work FDR is a central result in stochastic thermodynamics: it implies    that near equilibrium dissipation comes inevitably accompanied by fluctuations, and conversely that optimal protocols minimising dissipation will automatically also minimise fluctuations~\cite{Crooks2007,sivakThermodynamicMetricsOptimal2012}. The FDR also plays a crucial role in the estimation of  equilibrium free energy differences through measurements of non-equilibrium processes \cite{Hummer2001,Dellago2013}.

The main goal of this article is to experimentally validate the fundamental relation Eq. \eqref{eq:FDR} by using an electronic system.
For that, we consider one of the most relevant thermodynamic processes at the nanoscale: the conversion of information into work.  
Inspired by the  Szilard engine, we implement an information engine which uses one bit of information (an extra electron being located on the quantum dot or not) to extract work $W_{\rm ex} \equiv -W$
from a thermal reservoir at temperature~$T$. This cycle and its inverse, the so-called Landauer erasure, has been extensively studied in the literature, both theoretically~\cite{landauer_irreversibility_1961,landauer_dissipation_1988,sagawa_second_2008,sagawa_generalized_2010,horowitz_thermodynamic_2011,mandal_work_2012,barato_autonomous_2013,deffner_information_2013,parrondo_thermodynamics_2015,Faist2018,Miller2020,Proesmans2020} and experimentally~\cite{berut_experimental_2012,serreli_molecular_2007,raizen_comprehensive_2009,toyabe_experimental_2010,koski_experimental_2014,koski_-chip_2015}. 
It is well known that the average extractable work 
 is bounded  by 
$\langle W_{\rm ex} \rangle \leq  k_BT \ln 2$,
which can only be reached for infinitesimally slow processes --the so-called Landauer limit. 
Previous  experimental influential works have successfully managed to approach this limit to a high degree in various platforms at the level of average work \cite{berut_experimental_2012,serreli_molecular_2007,raizen_comprehensive_2009,toyabe_experimental_2010,koski_experimental_2014,koski_-chip_2015}.
In contrast, here  we  focus our attention on the behaviour of work fluctuations, and in particular  show how Eq. \eqref{eq:FDR} is satisfied as the Landauer limit is approached. 
As predicted from the FDR, our results clearly demonstrate a decrease in work ﬂuctuations as the dissipation is minimised  by increasing the operation cycle time of the engine. 
Therefore, the information about the electron's position can be exploited to extract work in an almost-deterministic fashion. To contrast this information-to-work process, we  consider a cyclic process where no information on the electron's position is available. In this case, we also observe that both work fluctuations and dissipation decrease according to \eqref{eq:FDR} as the operating time of the cycle increases. However, the average extracted work 
 always remains negative, $\langle W_{\rm ext} \rangle \leq 0$ and approaches zero for long cycle times, in agreement with the second law of thermodynamics. This contrasts the information engine that provides a positive finite work output with fluctuations vanishing towards long cycle times.
 
Our device is based on a semiconductor quantum dot which allows for the manipulation of discrete energy levels as opposed to the metallic system used in Ref~\citealp{koski_experimental_2014}. 
This work thus provides an experimental insight into the nature of thermodynamic ﬂuctuations in nanoscale systems as well as an alternative platform to study information-to-work conversion  with high efficiency.

\begin{figure}
    \centering
    \includegraphics[width=\columnwidth]{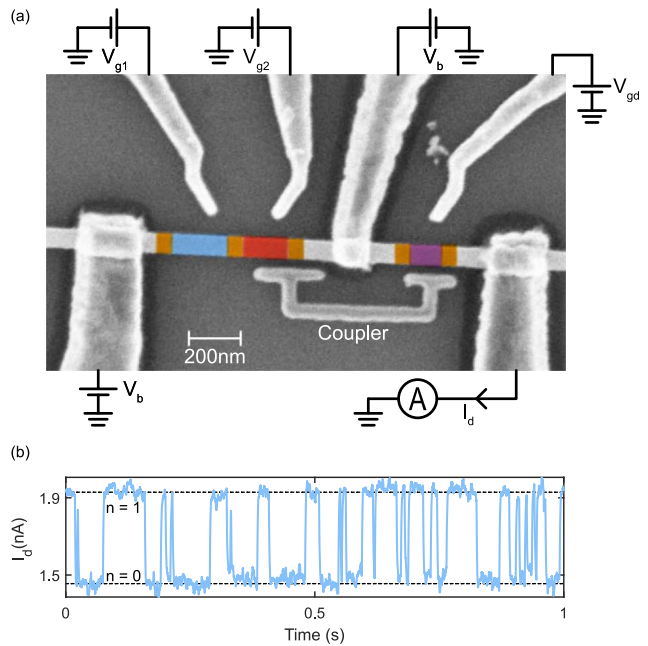}
    \caption{a) Scanning electron microscope image of the nanowire device. Embedded in the nanowire are three QDs, each aligned to one of the plunger gates $V_{g1}$,$V_{g2}$ or $V_{gd}$. Contacts separate the device into one part with two QDs and one part with a single QD. The coupler couples the two systems together, allowing the current $I_d$ through the lone QD to provide a measure of the charge state of the other system. Here, the QD involved in the experiment is marked in blue (close to the plunger gate with $V_{g1}$) while the quantum dot marked in red is put in Coulomb blockade. The sensor QD is marked in purple and the tunnel barriers are coloured orange. b) An example of the sensor current $I_d$ at the start of the experiment. $V_{gd}$ is set so that a high current represents the state $n=1$ and $V_{g1}$ is set so that the average $\langle n \rangle = 0.5$.}
    \label{fig:device}
\end{figure}

\emph{Set-up}. Our device is shown in Figure~\ref{fig:device}a). It consists of a quantum dot (QD) system formed with polytypes in an InAs nanowire~\cite{lehmann_general_2013,nilsson_transport_2016,chen_conduction_2017,barker_individually_2019}. The bit of information we use in our experiment is encoded in the occupancy of an electronic state in the leftmost QD (QD1). The state is at discrete energy level $E$ which we drive using the plunger gate voltage $V_{g1}$ with the lever arm  $\alpha = 1.7\times10^4k_BT/\mathrm{V}$. The rightmost QD is voltage biased with $V_b$ and used as an electrometer which detects the occupancy $n$ of the state~\cite{vandersypen_realtime_2004,gustavsson_electron_2009} through the current $I_d$ as shown in Figure~\ref{fig:device}b). Meanwhile, the middle QD is put in Coulomb blockade allowing tunneling from QD1 only to the left electronic reservoir. As in Refs.~\citealp{koski_experimental_2014, hofmann_heat_dissipation_2017, hofmann_equilibrium_2018}, our experiment is performed in a regime where $n \in \{0,1\}$ (i.e., the QD1 has an extra electron or not). $n$ switches whenever an electron tunnels between QD1 and the left-most fermionic reservoir at temperature $T=\SI{100}{\milli\kelvin}$, given by the cryostat temperature. The tunneling occurs with the rates $\Gamma_{\mathrm{in}}=\Gamma_0(1+bE)f(E)$ and $\Gamma_{\mathrm{out}}=2\Gamma_0(1+bE)(1-f(E))$ where $\Gamma_0 = \SI{64}{\hertz}$ and $b=0.025/k_BT$ were determined using a feedback protocol developed in Ref~\citealp{hofmann_measuring_2016}, and $f(E)=1/(1+e^{\beta(E)})$ is the Fermi-Dirac distribution. The average occupancy $p(t)$ of the QD state is then governed by the rate equation:
\begin{align}
    \dot{p}(t) =\Gamma_{\rm in} (1-p(t)) - \Gamma_{\rm out}\: p(t). 
    \label{eq: master equation}
\end{align}

\emph{Finite-time thermodynamics and work fluctuation-dissipation relation.} 
In the experiment, we consider protocols where the energy $E(t)$ is driven linearly from $E(0)\equiv E_i$ to $E(\tau)\equiv E_f$ for a time $\tau$. Work $W$ is performed on the system whenever $n(t)=1$ during the drive, i.e., $W \equiv \int_0^\tau {\rm d} t \hspace{1mm} n(t) \dot{E}(t)$. Note that work is a stochastic quantity due to fluctuations in $n(t)$ \cite{Jarzynski2011,Seifert_2012,Esposito2009,Campisi2011c,Goold_2016}. 
Sufficient repetitions of the protocol enables us to obtain the work probability distribution $P(W)$ experimentally. Its first moment is given by  $\langle W \rangle = \int_0^\tau {\rm d} t \hspace{1mm} p(t) \dot{E}(t)$, and we refer to the second moment as $\sigma_W^2$. 

The average work can be decomposed into a reversible and irreversible contribution: $\langle W \rangle = \Delta F + W_{\rm diss}$ \cite{Jarzynski2011,Seifert_2012,Esposito2009,Campisi2011c,Goold_2016}. Here, $\Delta F $ is the change of equilibrium free energy:
\begin{align}
    \Delta F \equiv \int_0^\tau {\rm d} t \hspace{1mm} p^{\rm eq}(t) \dot{E}(t) = \frac{1}{\beta} \ln \left(\frac{\mathcal{Z}_i}{\mathcal{Z}_f} \right),
\end{align}
where $p^{\rm eq}(t)= 1/(1+2e^{\beta E(t)})$ is the instantaneous equilibrium state, and $\mathcal{Z}_x=1+2e^{-\beta E_x}$ with $x=i,f$. The dissipated work $W_{\rm diss}$ is instead given by (see Appendix for a detailed derivation):
\begin{align}
   \hspace{-2.mm} W_{\rm diss}\hspace{-0.5mm} =\hspace{-1mm} \beta  \hspace{-0.5mm}\int_0^\tau {\rm d}t \hspace{1mm} [1-p^{\rm eq}(t)]p^{\rm eq}(t) \frac{\dot{E}(t)^2}{ \Gamma(t)} +\mathcal{O}\left(\frac{1}{\tau \Gamma_0} \right)^2 
   \label{eq:wdissexp}
\end{align}
where we have introduced $\Gamma(t)=\Gamma_0 (1+bE(t))$.  In Eq.~\eqref{eq:wdissexp} we have neglected contributions of order $(\tau \Gamma_0)^{-2}$, which is well justified in the experiment for $\tau \gtrsim 1$ s (for $\tau=1$ s we have: $1/\tau\Gamma_0 \approx 0.16$). 
On the other hand, in the Appendix, we show  that work fluctuations satisfy: $\sigma_w^2= 2 W_{\rm diss}/\beta  + \mathcal{O}(\Gamma_0\tau)^{-2} $. That is,  the FDR is satisfied at order $1/\tau\Gamma_0$, which is in agreement with the more general considerations of Refs.~\citealp{Mandal2016} and~\citealp{Miller2019work}. The regime where $W_{\rm diss}$ (and $\sigma_w^2$) decay with $1/\tau$ is often known in the literature as  the  low-dissipation regime, see e.g. Refs.~\citealp{Andresen1984,Esposito2010,Hernndez2015,Holubec2016,Ma2018,Ma2020}. 
\begin{figure}[t]
    \centering
    \includegraphics[width=\columnwidth]{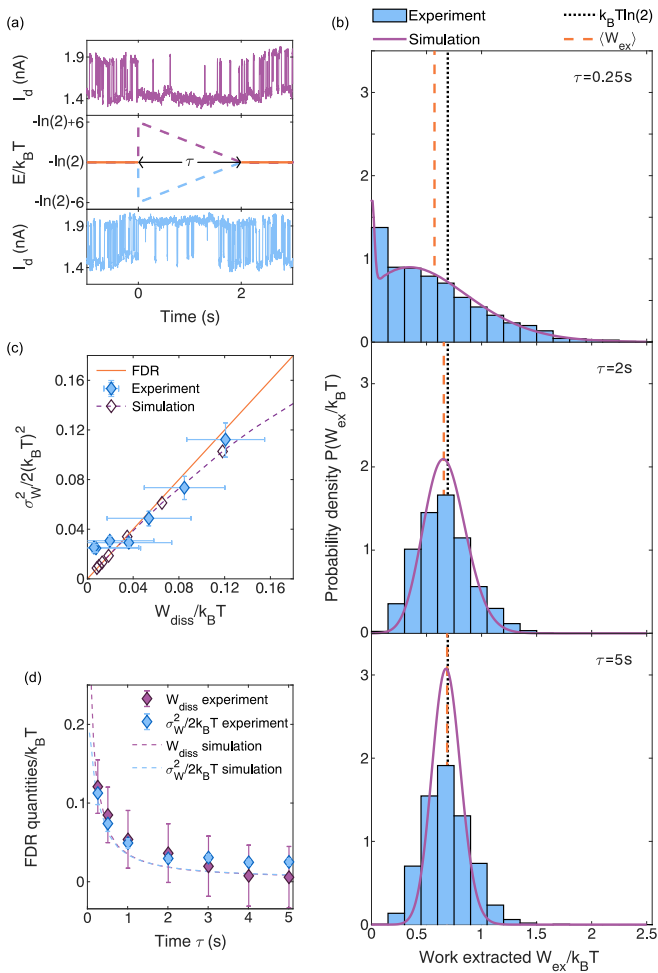}
    \caption{a)~The middle panel illustrates how the energy level is driven during the information-to-work conversion. Originally, it is offset from the reservoir Fermi level by $E = -k_BT\ln2$. At time 0, a measurement is made. Depending on the outcome, the energy is rapidly driven either up or down by $6k_BT$ before slowly being driven back to the starting point. In this example, the protocol time $\tau=\SI{2}{s}$. The upper (lower) panel shows the detector signal during one realization of the protocol where the energy was initially driven up (down). b)~The distributions of extracted work for the Szilard engine operation for the protocol lengths $\tau=$ \SI{0.25}{\second} (top), \SI{2}{\second} (middle) and \SI{5}{\second} (bottom) together with the theoretically expected distribution (purple).  The dotted black line indicates the Landauer limit $k_BT\ln 2$ and the orange dashed line indicates the mean extracted work. c) The two quantities from Eq.~\eqref{eq:FDR} plotted against each other. Here, $W_{\mathrm{diss}}=k_BT\ln 2-\langle W\rangle$. The FDR line is Eq.~\eqref{eq:FDR},
    whereas the simulations are obtained by a full counting statistics method with Eq.~\eqref{eq: master equation}.  d) Same quantities as in c), but  plotted against  $\tau$.}
    \label{fig:infoconvpos}
\end{figure}

\emph{Information-to-work conversion}. The device is operated as a Szilard engine and extracts work using the protocol developed in Ref.~\citealp{koski_experimental_2014} and shown in Figure~\ref{fig:infoconvpos}a). The protocol consists of three steps:
\begin{enumerate}
\item The QD is set to the condition of Figure~\ref{fig:device}b) where there is an equal chance to have $n=0$ and $n=1$, which occurs for a doubly spin-degenerate energy level when $E = -k_BT\ln 2$. 
\item A measurement of $n$ is made. This provides one bit of information, which is used to extract work. 
\item Depending on the measurement outcome, one of two feedback cycles is applied using $V_{g1}$ (the top and bottom panels of Figure~\ref{fig:infoconvpos}a) show example realizations of the two choices):
\begin{enumerate}
    \item If $n=0$, the level is rapidly raised by amplitude~$A = +6k_BT$. This process does no work to the system as the energy level is empty. The energy level is then slowly ramped down to the starting point for a time~$\tau$, extracting at most $\langle W_{\mathrm{ex}}\rangle = k_BT\ln 2$.
    \item If $n=1$, the level is instead lowered by the drive amplitude, extracting~$6k_BT$ of work and locking the state in $n=1$. While ramping the energy back up, on average $\langle W\rangle \geq 6k_BT-k_BT\ln 2$ of work is performed to the system.
\end{enumerate}

\end{enumerate}
 In both cases, the total amount of average work extracted in the cycle is bounded by $\langle W_{\rm ex} \rangle \leq k_B T \ln 2$. This limit can only be achieved  in the ideal reversible limit, but  finite-time implementations lead to both dissipation $W_{\rm diss}$ and fluctuations $\sigma_W^2$ as seen in the experimental data and supported by theory:
\begin{figure}[h]
    \centering
    \includegraphics[width=\columnwidth]{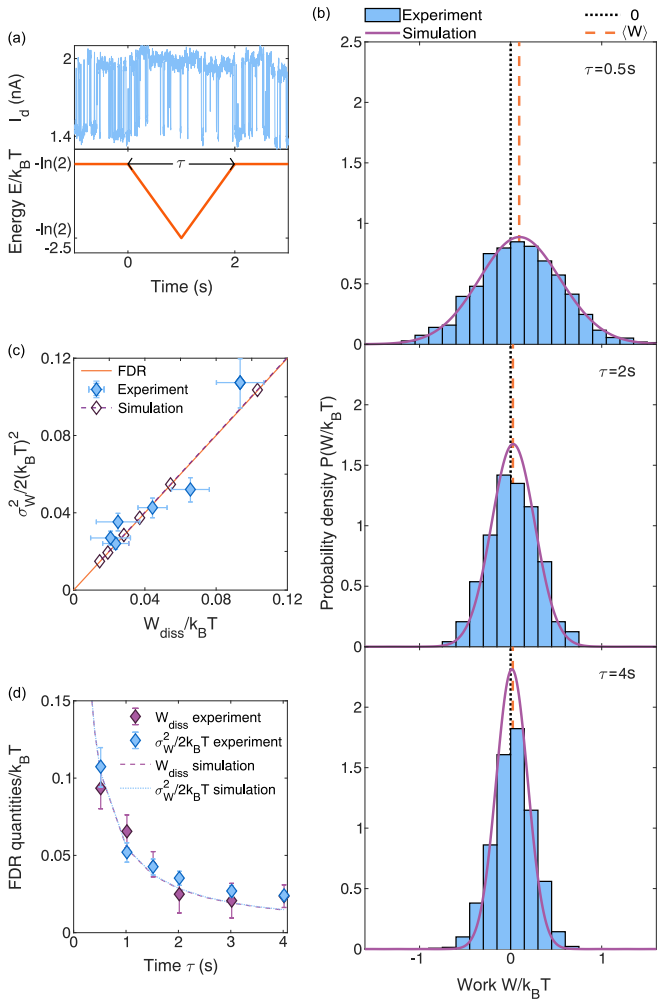}
    \caption{a) Top panel: The detector signal $I_d$ during the dissipative drive protocol, which is shown in the bottom panel. b) The work distribution of the $2.5k_BT$ dissipative drive for total protocol lengths \SI{0.5}{\second} (top), \SI{2}{\second} and \SI{4}{\second} (bottom) together with what is expected theoretically (purple). The orange line is the mean of the distribution which in this case is equal to $W_{\mathrm{diss}}$. The black line is drawn at $W = 0$ to help guide the eye. c) The two sides of Eq.~\eqref{eq:FDR} plotted against each other, for protocol lengths between \SI{0.5}{\second} and \SI{4}{\second}. The orange diagonal is given by  expression \eqref{eq:wdissexp}, which is obtained analytically in the slow-driving limit, and from which the FDR directly follows (see details in Appendix). d) The same quantities as in c), but now plotted against the protocol length.}
    \label{fig:chris25}
\end{figure}

We consider the performance of the information-to-work protocol for different times $\tau$, which we vary between \SI{0.25}{\second} and \SI{5}{\second}. Figure~\ref{fig:infoconvpos}b) shows how the distribution of extracted work changes as $\tau$ is increased. In the short time limit, there is a large peak at $W_{\mathrm{ex}} = 0$ which arises from instances where no tunneling events had time to occur. As the drive is made slower, these cases become less likely and the distribution starts forming a Gaussian distribution approaching the Landauer limit $k_BT\ln 2$. 
As we approach this limit, the distribution also sharpens, and the work FDR also becomes valid, as demonstrated in Figs.~\ref{fig:infoconvpos}c)-d) \footnote{An important remark is that in order to study the work FDR, we focus only on instances where the correct feedback drive was applied. There is a possibility that an electron tunnels between the measurement is made and the feedback is applied, which gives rise to a second distribution centered around $k_BT(\ln 2-6)$. In this particular experiment, the error rate was roughly $3\%$, but we expect to see fewer errors if the device is run in a regime with slower tunnelling rates.}: 
 Figure~\ref{fig:infoconvpos}c) shows the quantities of each side of Eq.~\eqref{eq:FDR} plotted against each other. The points line up with the diagonal, indicating that the work FDR does indeed hold.
 Furthermore, as seen in Figure~\ref{fig:infoconvpos}d), both  $\sigma_W^2$ and $W_{\mathrm{diss}}$ follow a $1/\tau$-dependence given by the analytical expression Eq. \eqref{eq:wdissexp}. 
 In general, we observe  very good agreement between theory and simulations. However, with largest $\tau$'s, $\sigma_W^2$ saturates in the experiment. This arises from offset charge drifts nearby QD1 that we observed to change at these time scales.
 
We now contrast the information-to-work conversion results with  a protocol where no information is involved. The protocol is presented in Figure~\ref{fig:chris25}a). Like the Szilard engine case, it also starts at $E = -k_BT \ln 2$ with a 50\% chance of QD1 being occupied. However, instead of making a measurement and quickly driving the level, it is slowly ramped down by the drive amplitude $2.5k_BT$ and then slowly ramped back up. On average,  this protocol only creates dissipation since the initial and final states are the same, hence $\Delta F=0$ in Eq. \eqref{eq:Wdiss}, which is why we refer to it as the dissipative drive. In this case, $W_{\mathrm{diss}} = \langle W \rangle$. The drive period $\tau$ is varied between \SI{0.5}{\second} and \SI{4}{\second}, and Figure~\ref{fig:chris25}b) shows the work distribution in those limits as well as an intermediate case. It is clear that as the drive slows down, $\langle W\rangle\rightarrow0$, while the distribution also sharpens up, in accordance with the FDR. The validity of the FDR for this cycle is shown in Figs.~\ref{fig:chris25}c-d), again finding very good agreement with experiment and theory expectations.
\begin{figure}
    \centering
    \includegraphics[width = \columnwidth]{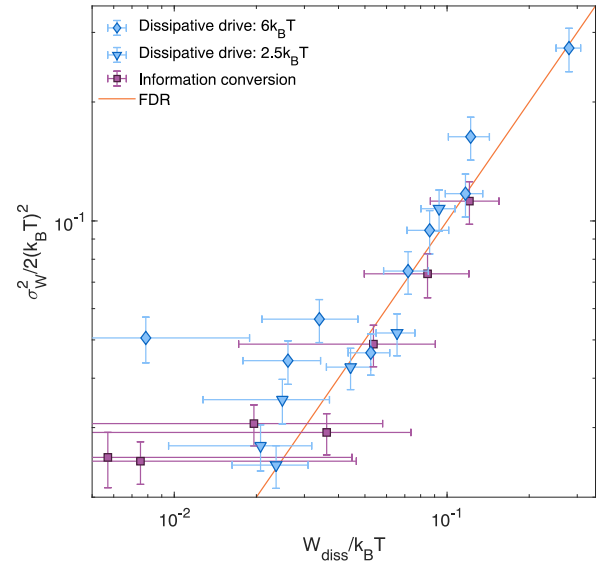}
    \caption{The quantities in the work FDR in equation~\ref{eq:FDR} plotted against each other for the information conversion protocol (purple) and the two dissipative protocols (blue, $6k_BT$ as diamonds and $2k_BT$ as triangles). To make the points easier to visually separate, the axes are logarithmic. }
    \label{fig:alldrives}
\end{figure}
Figure~\ref{fig:alldrives} summarizes the results for all protocols tested. This includes the previously discussed information-to-energy conversion drive and the $2.5k_BT$ dissipative drive as well as an additional dissipative drive with a $6k_BT$ amplitude. All points follow Eq.~\eqref{eq:FDR}, except the longest drives where the fluctuations saturate due to the offset charge drifts. These results encapsulate the main finding of this article: the experimental verification of the fundamental thermodynamic relation Eq.~\eqref{eq:FDR} at the level of a single-electron quantum dot.

\emph{Conclusions}. In summary, we experimentally verified the work FDR for an information-to-work  engine, implemented in a semiconductor quantum dot system.  In agreement with the FDR, we observed a sharpening of the work distribution as the Landauer or reversible limit  is approached.
This implies that  $k_BT \ln 2$ can be extracted almost \emph{deterministically} from one bit of information (the electron's position in the dot). We contrasted these results with a process where no information about the electron was used. In this case, we also observed a decrease of dissipation accompanied by a decrease in fluctuations, but no work could be extracted on average.

Overall, our results highlight that the fundamental thermodynamic relation Eq. \eqref{eq:FDR} is valid at the level of a single electron. 
These measurements constitute the first steps of studying information-to-work conversion using semiconductor quantum dots. Similar devices could be used to study, for instance, optimal protocols for information-to-work conversion~\cite{scandi_thermodynamic_2019,Abiuso2020entropy,Proesmans2020,Miller2020b,zhen2021universal,konopik2021fundamental},  as well as thermodynamic uncertainty relations~\cite{potts_thermodynamic_2019}. Another future direction of interest is to, instead of inferring work output from the QD occupation, use information to generate a current and measure the work output directly~\cite{annby-andersson_maxwells_2020}. Finally, a longer-term challenge is to experimentally show violations of the work FDR Eq. \eqref{eq:FDR}  due to the presence of quantum coherence, as theoretically predicted in Ref.~\citealp{Miller2019work}.

\emph{Acknowledgments.} We would like to thank Christopher Jarzynski for helpful discussions. D. B. and V. F. M. thank for financial support from NanoLund, Swedish Research Council, Foundational Questions Institute and the Knut and Alice Wallenberg Foundation (KAW) via Project No. 2016.0089. M. P.-L.  acknowledges funding from Swiss National Science Foundation through an Ambizione grant PZ00P2-186067. M. S. acknowledges support from the European Union’s Horizon 2020 research and innovation programme under the Marie Sk\l{}odowska-Curie grant agreement No 713729, and from the Government of Spain (FIS2020-TRANQI and Severo Ochoa CEX2019-000910- S), Fundacio Cellex, Fundaci\'{o} Mir-Puig, Generalitat de Catalunya (SGR 1381 and CERCA Programme).

\bibliography{FDRbib}

\newpage
\appendix
\onecolumngrid

\section{Derivation of the FDR}

For completeness, here we add a proof of the work FDR for the driven quantum dot. The results follow from the general considerations of Refs.~\citealp{Mandal2016,Miller2019work}, but provide a more explicit derivation for the specific setting considered. 

Recall that we consider a quantum dot with a two-fold degenerate energy level $E(t)$ that is driven for a time $t \in [0,\tau]$, which is in contact with a thermal fermionic reservoir. 
It is convenient to express the quantities of interest as $x_s \equiv x(s\tau)$ with $s\in (0,1)$. In terms of the renormalised variables, the equation of motion reads:
\begin{align}
    \frac{1}{\tau} \dot{p}_s = \Gamma_s (p^{\rm eq}_s -p_s)
    \label{difeqapp}
\end{align}
with $\dot{q}_s={\rm d} q_s/{\rm d} s$, and where we introduced $\Gamma_s=\Gamma_0 (1+b E_s)$ and $p^{\rm eq}_s=1/(1+2e^{\beta(E_s)})$. 

The stochastic variable   $n_s \in \{0,1\}$ determines whether the dot is occupied or not at time $s\tau$. Hence, $\mathcal{E}[n_s]=p_s$ where $\mathcal{E}[.]$ denotes averaging over all trajectories. The  work extracted  in a given trajectory reads:
$W=\int_0^\tau {\rm d} s \hspace{1mm} n_s \dot{e}_s$. The  average work is given by 
\begin{align}
    \langle W \rangle=
\mathcal{E}[ W ] =\int_0^1 {\rm d} s \hspace{1mm} \mathcal{E}[n_s] \dot{E}_s =\int_0^1 {\rm d} s \hspace{1mm} p_s \dot{E}_s.
\label{AvWapp}
\end{align} 
On the other hand, the work fluctuations read
\begin{align}
\sigma^2_W=\mathcal{E}[ W^2 ]-(\mathcal{E}[W ])^2
\label{workFluctapp}
\end{align}
with 
$\mathcal{E}[ W^2 ] =\int_0^1 {\rm d} s \hspace{1mm}\int_0^1 {\rm d} t \hspace{1mm}  \mathcal{E}[n_s n_t] \dot{E}_s \dot{E}_t$.
For $t\geq s$, $\mathcal{E}[n_s n_t]$ is given by 
$\mathcal{E}[n_s n_t]=p_s p_{t|s}$, where  $p_{t|s}$ is the conditional probability of having $n_t=1$ given that $n_s =1 $ at time $s<t$ (this ensures that the product $n_t n_s$ is non-zero).
Hence we have
$\mathcal{E}[W^2 ] =2 \int_0^\tau {\rm d} t \hspace{1mm}\int_0^t {\rm d} s \hspace{1mm} p_s p_{t|s} \dot{E}_s \dot{E}_t$, and also:
\begin{align}
\sigma_W^2 = 2 \int_0^\tau {\rm d} t \hspace{1mm}\int_0^t {\rm d} s \hspace{1mm} p_s (p_{t|s} - p_t) \dot{E}_s \dot{E}_t
\label{ex2}
\end{align}
for the work fluctuations. 
In principle, both $ \langle W \rangle$ and $\sigma^2_W$ can be computed exactly by solving the differential equation~\eqref{difeqapp}. We obtain the formal solutions:
\begin{align}
    p_s = p_0 e^{-\tau \int_0^s \Gamma_y \hspace{1mm}{\rm d}y}+ \tau \int_0^s e^{-\tau \int_x^s \Gamma_y \hspace{1mm}{\rm d}y} \Gamma_x  p^{\rm eq}_x \hspace{1mm}{\rm d}x.
\end{align}
and
\begin{align}
    p_{t|s} =  e^{-\tau \int_s^t \Gamma_y \hspace{1mm}{\rm d}y}+ \tau \int_s^t e^{-\tau \int_x^t \Gamma_y \hspace{1mm}{\rm d}y} \Gamma_x  p^{\rm eq}_x \hspace{1mm}{\rm d}x.
    \label{formSolptsApp}
\end{align}

In order to obtain the FDR, we are interested in obtaining $ \langle W \rangle$ and $\sigma^2_W$ analytically in the slow driving regime, i.e., at leading order in $1/\tau$. We proceed differently for each term. For  $ \langle W \rangle$, it is convenient to solve the differential equation \eqref{difeqapp}  
 perturbatively for $1/\tau$ by introducing the expansion:
    $p_s=p_s^{\rm (0)}+ \frac{1}{ \tau}p_s^{\rm (1)}+...$ \cite{cavinaSlowDynamicsThermodynamics2017} (see also  Refs. \citealp{Campisi2012geometric,Deffner2013,Ludovico2016,cavinaSlowDynamicsThermodynamics2017,Abiuso2020entropy}).
At zeroth order ($1/\tau =0$), we simply find the equilibrium solution $p_s^{(0)}=p_s^{\rm eq}$. At first order, we have $p_s^{(1)}= - \dot{p}_s^{(0)} /\Gamma_s$, and hence:
\begin{align}
    p_s&=p_s^{\rm eq}- \frac{1}{\tau \Gamma_s} \dot{p}_s^{\rm eq}+\mathcal{O}\left(\frac{1}{\tau \Gamma_0} \right)^2
    \nonumber\\
    &=p_s^{\rm eq}+ \frac{\beta \dot{E}_s}{\tau \Gamma_s} (1-p_s^{\rm eq})p_s^{\rm eq}+\mathcal{O}\left(\frac{1}{\tau \Gamma_0} \right)^2
     \label{pertSolp}
\end{align}
where we used 
$p^{\rm eq}_s=1/(1+2e^{\beta(E_s)})$ in the second line. Plugging this solution into \eqref{AvWapp}, we obtain:
\begin{align}
    \langle W \rangle = \Delta F + \frac{\beta }{\tau} \int_0^1 {\rm d}s \hspace{1mm} (1-p_s^{\rm eq})p_s^{\rm eq} \frac{\dot{E}_s^2}{ \Gamma_s} +\mathcal{O}\left(\frac{1}{\tau \Gamma_0} \right)^2
\end{align}
with $\Delta F = \int_0^1 {\rm d} s \hspace{1mm} p_s^{\rm eq} \dot{E}_s$. Obtaining $\sigma_W^2$ at leading order in $1/\tau$ is more subtle, and we have to proceed differently. The underlying reason is that the perturbative solution \eqref{pertSolp}  disregards the memory of initial condition (which decays exponentially with time). This is well justified for $p_s$  away from  $s=0$, but not for  $p_{t|s}$ where the regime $t\approx s$ is the most relevant one. Instead, we can expand directly the formal solution \eqref{formSolptsApp} around $s=t$, obtaining at leading order:
\begin{align}
    p_{t|s} &= e^{-\tau \Gamma_t (t-s)}+\tau \int_s^t e^{-\tau \Gamma_t (t-x)} \Gamma_t p_t^{\rm eq} {\rm d}x +\mathcal{O}\left((t-s)e^{-\tau \Gamma_0 (t-s)}\right)
    \nonumber\\
    &= p_t^{\rm eq} +e^{-\tau \Gamma_t (t-s)} (1-p_t^{\rm eq})+ \mathcal{O}\left((t-s)e^{-\tau \Gamma_0 (t-s)}\right)
\end{align}
Plugging this expression into \eqref{ex2}, and consistently expanding around $s=t$, we obtain:
\begin{align}
\sigma_W^2 &= 2 \int_0^\tau {\rm d} t \hspace{1mm} p_t^{\rm eq} (1-p_t^{\rm eq}) \dot{E}_t^2 \int_0^t \left[ e^{-\tau \Gamma_t (t-s)}+ \mathcal{O} \left((t-s)e^{-\tau \Gamma_0 (t-s)}\right) \right]{\rm d} s
§\nonumber\\
&= \frac{2}{\tau} \int_0^\tau {\rm d} t \hspace{1mm} p_t^{\rm eq} (1-p_t^{\rm eq}) \frac{\dot{E}_t^2}{\Gamma_t}  \hspace{1mm} +\mathcal{O}\left(\frac{1}{\tau \Gamma_0} \right)^2.
\end{align}
This provides the desired result:
\begin{align}
    W_{\rm diss} = \frac{\beta}{2} \sigma_W^2   +\mathcal{O}\left(\frac{1}{\tau \Gamma_0} \right)^2.
\end{align}
\end{document}